\documentclass[aps,preprint,tightenlines,showpacs,showkeys]{revtex4-1}
\usepackage[latin9]{inputenc}
\usepackage{amsmath}
\usepackage{amssymb}
\usepackage{esint}
\usepackage{graphicx}
\usepackage{float}

\makeatletter

\usepackage{dcolumn}
\usepackage{bm}

\makeatother

\begin{document}

\title{$q$-Deformed Einstein equations from entropic force}
\author{Mustafa Senay$^{1}$\footnote{E-mail: mustafasenay86@hotmail.com} and Salih Kibaro\u{g}lu$^{1}$\footnote{E-mail: salihkibaroglu@gmail.com}}
\date{\today}
\begin{abstract}
In this study, we investigate the influences of fermionic $q$-deformation
on the Einstein equations by taking into account of Verlinde's entropic
gravity approach and Strominger's proposal on quantum black holes. According to Verlinde's proposal, gravity is interpreted as an entropic force. Moreover,
Strominger's suggetion claims that extremal black holes obey deformed statistics
instead of the standard Bose or Fermi statistics. Inspired by Verlinde's
and Strominger's suggestions, we represent some thermostatistical
functions of VPJC-type $q$-deformed fermion gas model for the high-temperature
limit. Applying the Verlinde's entropic gravity approach to the $q$-deformed
entropy function, $q$-deformed Einstein equations with the effective cosmological
constant are derived. The results obtained in this work are compared
with the related works in the literature.
\end{abstract}
\affiliation{$^{1}$Department of Basic Sciences, Naval Academy, National Defence University, 34940 \.{I}stanbul,
Turkey }
\maketitle

\section{Introduction}

There is a long story behind researching the law of gravity from ancient
times to now. Until the beginning of the 20th century, Newton's law
of gravitation was accepted as an almost completed theory. But the
Einstein's theory of gravitation which is well known as \textquotedblleft general
theory of relativity\textquotedblright{} emerged as the most successful theory
until that time. After this important study, more complicated theories were
developed such as gauge theory of gravity, quantum gravity, and string
theory. 

Recent years, the thermodynamic approximation of gravity has been
underdeveloped. One can say that this aspect is based on Hawking and
Bekenstein's papers \cite{hawking1971,hawking1975,bekenstein1973}
which are about the black hole physics and thermodynamics. Hawking
showed us in his paper \cite{hawking1971} that the black hole area
cannot diminish in any process. From this idea, Bekenstein established
the relationships between the black hole surface area and its entropy.

In 1995 Jacobson \cite{jacobson1995} used the black hole entropy
to develop a gravity theory based on thermodynamics. He claimed that
the theory of gravitation can be expressed by the first law of thermodynamics on a local
Rindler horizon, not as a fundamental interaction. He also found the
null-null components of the Einstein equation with the help of this thermodynamical approximation. Another important study
was proposed by Verlinde \cite{verlinde2011} who theorized the emergence
of gravity as an entropic force by using the holographic viewpoint
of gravity \cite{mann2011}. This method which is called as entropic
gravity provide him to find Newton's second law of gravity and Verlinde also found
the time-time components of Einstein equations by using Tolman-Komar mass and the equipartition rule.  Likewise, all the
components of Einstein equations were derived in Ref. \cite{li2012}.
More detailed review can be found in Ref. \cite{padmanabhan2010}.

Discovery of entropic origin of gravity gives us a new method to modify
Newton's law of gravitation or the general theory of relativity by
modification of statistical thermodynamics. By means of this idea, many authors have extensively investigated the unification of thermodynamics and gravity theories such as Milgrom's
MOND theory \cite{sheykhi2012}, modified dark matter (MDM) \cite{ho2010,ng2013,edmonds2014},
modification of Coulomb law \cite{hendi2012}, and extremal black
holes \cite{Dil2015,dil2016B,dil2017}.

Moreover, according to the Strominger's idea in Ref. \cite{Strominger1993},
black holes can be considered as a \textit{q}-deformed bosons or fermions
and their statistics can be explained by \textit{q}-deformed bosons
and fermions statistics. Recently, several authors have intensively
studied the thermodynamical and statistical properties of \textit{q}-deformed
bosons and fermions \cite{Gavrilik2011,Gavrilik2013,Algin2013,Algin2015,Cai2007,Mirza2011,Lavagno2000,Lavagno2002,Lavagno2010,Marinho2012,Zeng2012,NarayanaSwamy2006a,NarayanaSwamy2006b}.
Also, the \textit{q}-deformed systems have wide-ranging applications in
many areas such as understanding higher-order effects in the many-body
interactions \cite{Sviratcheva2004}, vibrational spectra of diatomic
molecules \cite{Bonatsos1992}, vortices in superfluid films \cite{Bonatsos1996},
and phonon spectrum in $^{4}He$ \cite{Monterio1996}. Another application
is that black holes were considered as \textit{q}-deformed bosons
and thus \textit{q}-deformed Einstein field equations were obtained
\cite{Dil2015}.

In the view of the above motivations, in this work, black
holes are considered as \textit{q}-deformed fermions which are known as Visvanathan-Parthasarthy-Jagannathan-Chaichian (VPJC) in the literature. Historically, the fermionic oscillator algebra of this model was presented
in \cite{Parthasarathy1991,Viswanathan1992} and some of its statistical
properties were investigated in \cite{Chaichian1993}.
Therefore, this model was called as VPJC-type \textit{q}-deformed
fermion model in \cite{Algin2011}. In addition, the high-temperature
thermostatistical properties of this model for three spatial dimension
were extensively analyzed in \cite{Algin2012}. Using the deformed
entropy function of this fermion model, the deformed gravitational
field equations of Einstein can be derived by taking account of Verlinde's
approach.

This paper is organised as follows: In Sec. II, a brief summary
of the VPJC-type \textit{q}-oscillators model is given and some thermostatistical
functions of the model for high-temperature limit are represented. In Sec. III, \textit{q}-deformed
Einstein field equations are derived with the effective cosmological constant. The conclusions
are presented in Sec. IV.

\section{High-temperature Thermodynamic of the $q$-Deformed fermion model}

In this section, we shall briefly review the essential relations of
\textit{q}-oscillators introduced in \cite{Parthasarathy1991,Viswanathan1992,Chaichian1993}. Also,
we re-examine some thermodynamical properties of VPJC-type fermion
model introduced in \cite{Algin2012} on high-temperature limit.

VPJC-type \textit{q-}deformed fermion oscillators algebra is defined
in terms of the deformed fermionic creation and annihilation operator  $f^{*}$,
$f$, and number operator $\hat{N}$ \cite{Viswanathan1992,Chaichian1993}:
\begin{eqnarray}
 & ff^{*}+qf^{*}f=1,\nonumber \\
 & [\hat{N},f^{*}]=f^{*},\,\,\,\,\,\,[\hat{N},f]=-f,
\end{eqnarray}
where \textit{q} is the real positive deformation parameter having
the range $q<1$. Furthermore, the deformed fermionic number operator
for the VPJC-fermion model is defined \cite{Algin2011} as
\begin{equation}
f^{*}f=[\hat{N}]=\frac{1-(-1)^{\hat{N}}q^{\hat{N}}}{1+q},
\end{equation}
whose spectrum is 
\begin{equation}
[n]=\frac{1-(-1)^{n}q^{n}}{1+q},
\end{equation}
which is the \textit{q}-fermionic basic number for the model. Moreover,
VPJC-type \textit{q}-deformed fermion model has the following fermionic
Jackson derivative (JD) operator
\begin{equation}
D_{x}^{(q)}f(x)=\frac{1}{x}\left[\frac{f(x)-f(-qx)}{1+q}\right],
\end{equation}
for any function $f(x)$. This fermionic JD operator does not reduce
to the standard derivative in the limit $q=1$ and is used to examine
thermodynamic properties of our model.

The VPJC-type \textit{q}-deformed fermion model has the following
mean occupation number \cite{Algin2011} as
\begin{equation}
n=\frac{1}{\left|\ln q\right|}\left|\ln\left(\frac{\left|z^{-1}e^{\beta\varepsilon}-1\right|}{z^{-1}e^{\beta\varepsilon}+q}\right)\right|,
\end{equation}
where $\beta=1/kT$ with \textit{k }being the Boltzmann constant and
$z=\exp(\mu/kT)$ is the fugacity. Using the relations in Eqs. (4)
and (5), the logarithm of the fermionic grand partition function of
the model can be found \cite{Algin2012} as
\begin{equation}
\ln Z=\frac{(1+q)}{\left|\ln q\right|}\sum_{i}\left|\ln\left|(1-ze^{-\beta\varepsilon})\right|\right|.
\end{equation}
For a large volume and a large number of particles, the sums over states
can be replaced with the integral. Accordingly, the equation of state
$PV/kT=\ln Z$ can be written as
\begin{equation}
\frac{P}{kT}=\frac{(1+q)}{\left|\ln q\right|}\frac{4\pi}{h^{3}}\intop_{0}^{\infty}dp\,p^{2}\left|\ln\left|(1-ze^{-\beta p^{2}/2m})\right|\right|,
\end{equation}
where $\varepsilon=p^{2}/2m$. Following the prescription of the fermionic
JD operator in Eq. (4), we may re-express the above equation as
\begin{equation}
\frac{P}{kT}=\frac{1}{\lambda^{3}}f_{5/2}(z,q),
\end{equation}
where $\lambda=h/\sqrt{2\pi mkT}$ is the thermal wavelength and \textit{q}-deformed
Fermi-Dirac function $f_{n}(z,q)$ is defined as
\noindent 
\begin{equation}
f_{n}(z,q)=\frac{1}{\left|\ln q\right|}\left[\sum_{l=1}^{\infty}(-1)^{(l-1)}\frac{(zq)^{l}}{l^{n+1}}-\sum_{l=1}^{\infty}\frac{z^{l}}{l^{n+1}}\right].
\end{equation}

In a similar manner, the particle density of the \textit{q}-deformed
fermion model can be obtained as
\begin{equation}
\frac{N}{V}=\frac{1}{\lambda^{3}}f_{3/2}(z,q).
\end{equation}
From the thermodynamic relation $U=-(\partial\ln Z/\partial\beta)$$_{z,V}$,
the internal energy of the model can be derived as
\begin{equation}
\frac{U}{V}=\frac{3}{2}\frac{kT}{\lambda^{3}}f_{5/2}(z,q).
\end{equation}
The Helmholtz free energy $A=\mu N-PV$ can be found from Eqs. (8)
and (10) as
\begin{equation}
A=\frac{kTV}{\lambda^{3}}\left[f_{3/2}(z,q)\ln z-f_{5/2}(z,q)\right].
\end{equation}
The \textit{q}-deformed entropy function of the fermion model can
be derived from the relation $S=(U-F)/T$ as
\begin{equation}
S=\frac{kV}{\lambda^{3}}\left[\frac{5}{2}f_{5/2}(z,q)-f_{3/2}(z,q)\,\ln z\right].
\end{equation}
If we put the one-particle kinetic energy \textit{E }instead of \textit{kT
}in Eq. (13), the deformed entropy function can be re-expressed as
\begin{equation}
S=\frac{(2\pi m)^{3/2}V}{Th^{3}}E^{5/2}\widetilde{F}(z,q),\label{eq: def. S}
\end{equation}
where
\begin{equation}
\widetilde{F}(z,q)=\frac{5}{2}f_{5/2}(z,q)-f_{3/2}(z,q)\ln z.
\end{equation}
By using the deformed entropy function in Eq. (14), the deformed Einstein
equations can be obtained. For this reason, we will apply the Verlinde's
approach to our deformed fermion model in the next section. 

\section{$q$-Deformed Einstein equations }

From Bekenstein \cite{bekenstein1973} to now many people have studied
the relation between statistical thermodynamics and gravity. They
have also revealed some close connection between mentioned fields. In 2011 the theory which is known as the Entropic Gravity is constructed by Verlinde \cite{verlinde2011}.
This theory shows that the gravity theory can be explained by the laws
of thermodynamics. From this point of view an interesting idea has
emerged; if one modifies the thermodynamic quantities such as
temperature or entropy, one can get modified theory of gravitation. In
the light of this idea, we would like to derive the Einstein equations
with cosmological constant by using $q$-deformed fermion gas model.
In this section, firstly, we briefly give Verlinde's approach on entropic
gravity, secondly, the deformed temperature will be obtained and in
the last modified Einstein equations will be found. 

Considering time like Killing vector $\xi^{a}$ with a static background,
we can relate the temperature and the entropy gradient of a system
by using the Killing vector field to see the effect of the equivalence
principle or the emergence of inertia. Now we will give some theoretical
background about force, entropy and temperature. In general relativity,
one can generalise the Newton's potential \cite{wald1984} as
\begin{equation}
\phi=\frac{1}{2}\log\left(-\xi^{a}\xi_{a}\right),\label{eq:Newton Pot.}
\end{equation}
also, its exponential corresponds to the redshift factor that relates
the local time coordinate system. Taking the four-velocity as $u^{a}$
and its acceleration $a^{b}=u^{a}\nabla_{a}u^{b}$ can be defined
by using the Killing vector field,
\begin{equation}
u^{b}=e^{-\phi}\xi^{b},
\end{equation}
\begin{equation}
a^{b}=e^{-2\phi}\xi^{a}\nabla_{a}\xi^{b}.
\end{equation}
Here the acceleration expression can be written as below by using the
Killing equation $\nabla_{a}\xi_{b}+\nabla_{b}\xi_{a}=0$ and the definition
of Newton's potential (\ref{eq:Newton Pot.}),
\begin{equation}
a^{b}=-\nabla^{b}\phi.
\end{equation}
These definitions provide us to write the force such as,
\begin{equation}
F_{a}=-me^{\phi}\nabla_{a}\phi,
\end{equation}
where the additional term $e^{\phi}$ is due to the redshift factor.
Verlinde considers the black hole horizon as a holographic screen
for constructing entropic gravity based on Bekenstein's idea which
is about relationships between black hole surface area and its entropy.
The change of entropy of a particle which is located very close to
the screen can be written as follows
\begin{equation}
\nabla_{a}S=-2\pi\frac{m}{\hbar}N_{a},\label{eq: cov S}
\end{equation}
where $N^{a}$ is unit outward pointing vector that normal to screen
and $m$ is mass of particle. We also know that the acceleration is
perpendicular to screen. So, the entropic force takes the following
form
\begin{equation}
F_{a}=T\nabla_{a}S=-me^{\phi}\nabla_{a}\phi.\label{eq:F T S}
\end{equation}
This equation can be seen as a relativistic analogue of the Newton's law
of inertia, where $T$ is the non-deformed temperature according to
the Verlinde's approach \cite{verlinde2011} given as
\begin{equation}
T=\frac{\hbar}{2\pi}N^{a}e^{\phi}\nabla_{a}\phi.
\end{equation}

Considering the relation (\ref{eq:F T S}), if we deform entropy or
temperature of the system, deformed force can be obtained or in the other
words, we can obtain generalised gravitational theory. In our condition
the entropy is thought as extremal \cite{verlinde2011} and it can
be written,
\begin{equation}
\frac{d}{dx^{a}}S(E,x^{a})=0.
\end{equation}
Also, one can obtain the following relation
\begin{equation}
\frac{\partial S}{\partial E}\frac{\partial E}{\partial x^{a}}+\frac{\partial S}{\partial x^{a}}=0,
\end{equation}
where $\frac{\partial E}{\partial x^{a}}=-F_{a}$ and $\frac{\partial S}{\partial x^{a}}=\nabla_{a}S$.
To obtain deformed temperature we will use the deformed entropy found
in Eq. (\ref{eq: def. S}) and the last relation. So we get
\begin{equation}
\frac{5(2\pi mE)^{3/2}}{2h^{3}}\widetilde{F}(z,q)F_{a}=T\nabla_{a}S.
\end{equation}
Using Eqs. (\ref{eq: cov S}) and (\ref{eq:F T S}) the deformed temperature
can be found as
\begin{equation}
T=\frac{5V}{8\sqrt{\pi}}\frac{\left(2mE\right)^{3/2}}{h^{2}}\widetilde{F}\left(z,q\right)N^{a}e^{\phi}\nabla_{a}\phi.\label{eq: def T}
\end{equation}

Now we would like to obtain deformed Einstein equations originated
from the entropic force by using deformed temperature in Eq. (\ref{eq: def T}).
According to Bekenstein, if there is a test particle near the black
hole horizon which is distant from a Compton wavelength, it increases black hole mass. This process is identified as
one bit of information. Considering the holographic screen on closed
surface of constant redshift, the differential of number of bits can
be written as
\begin{equation}
dN=\frac{dA}{G\hbar},\label{eq: bit}
\end{equation}
where $A$ is the area of closed surface on the screen and $G$ is
the Newton's gravitational constant. Assuming the total energy $E$
associated with the total mass $M$ distributed over all bits on the
screen. According to the equipartition law of energy, $E$ is equal
to $\frac{1}{2}TN$. Using the relation $E=M$ under the condition
$c=1$, the total mass can be written as
\begin{equation}
M=\frac{1}{2}\int_{s}TdN.
\end{equation}
Using Eqs. (\ref{eq: def T}) and (\ref{eq: bit}), the total mass
expression goes following form
\begin{equation}
M=\frac{1}{4\pi G}\frac{5V\left(2\pi mE\right)^{3/2}}{2h^{3}}\widetilde{F}\left(z,q\right)\intop_{s}e^{\phi}\nabla\phi dA.
\end{equation}
This equation corresponds to the Komar mass for our deformed system and
can be defined as $q$-deformed Gauss's law. Also, the Komar mass can be written in terms of Killing vector $\xi^{a}$
as
\begin{equation}
M=\frac{1}{4\pi G}\frac{5V\left(2\pi mE\right)^{3/2}}{2h^{3}}\widetilde{F}\left(z,q\right)\intop_{s}dx^{a}\wedge dx^{b}\epsilon_{abcd}\nabla^{c}\xi^{d}.
\end{equation}
Using the Stokes theorem, the Killing equation and the relation $\nabla_{a}\nabla^{a}\xi^{b}=-R_{\,\,\,a}^{b}\xi^{a}$,
we get the Komar mass as follows
\begin{equation}
M=\frac{\alpha\left(z,q\right)}{4\pi G}\int_{\Sigma}R_{ab}n^{a}\xi^{b}dV,\label{eq:mass1}
\end{equation}
where $R_{ab}$ is Ricci tensor and the factor $\alpha\left(z,q\right)$
is
\begin{equation}
\alpha\left(z,q\right)=\frac{5V\left(2\pi mE\right)^{3/2}}{2h^{3}}\widetilde{F}\left(z,q\right).
\end{equation}
Moreover, an alternative expression of the Komar mass can be given as
\cite{wald1984}
\begin{equation}
M=2\int_{\Sigma}\left(T_{ab}-\frac{1}{2}g_{ab}T+\frac{\Lambda}{8\pi G}g_{ab}\right)n^{a}\xi^{b}dV,\label{eq:komar em}
\end{equation}
where $T_{ab}$ is energy momentum tensor, $g_{ab}$ is (mostly minus) space-time
metric tensor and $\Lambda$ is the standard cosmological constant. After these
calculations, if we compare the Eqs. (\ref{eq:mass1}) and (\ref{eq:komar em})
then we get
\begin{equation}
\frac{\alpha\left(z,q\right)}{4\pi G}\int_{\Sigma}R_{ab}n^{a}\xi^{b}dV=2\int_{\Sigma}\left(T_{ab}-\frac{1}{2}g_{ab}T+\frac{\Lambda}{8\pi G}g_{ab}\right)n^{a}\xi^{b}dV,
\end{equation}
and the Ricci tensor which satisfies the Bianchi identity as $\nabla R_{ab}=0$, can be obtained as follows,
\begin{equation}
R_{ab}=8\pi G\alpha\left(z,q\right)^{-1}\left(T_{ab}-\frac{1}{2}g_{ab}T+\frac{\Lambda}{8\pi G}g_{ab}\right).\label{eq: R_ab}
\end{equation}
Finally, using the Eq. (\ref{eq: R_ab}) and trace of it, we get following equation,
\begin{equation}
R_{ab}-\frac{1}{2}g_{ab}R+\Lambda^{\left(q\right)} g_{ab}=8\pi G^{\left(q\right)}T_{ab}.\label{eq: dE_eq}
\end{equation}
where coefficients $G^{\left(q\right)}$ and $\Lambda^{\left(q\right)}$ are, respectively, called the effective gravitational constant and the effective cosmological constant defined as
\begin{equation}
G^{\left(q\right)}=\alpha\left(z,q\right)^{-1}G=\left(\frac{2}{5V\left(2\pi mE\right)^{3/2}}\frac{h^{3}}{\widetilde{F}\left(z,q\right)}\right)G,
\end{equation}\begin{equation}
\Lambda^{\left(q\right)}=\alpha\left(z,q\right)^{-1}\Lambda=\left(\frac{2}{5V\left(2\pi mE\right)^{3/2}}\frac{h^{3}}{\widetilde{F}\left(z,q\right)}\right)\Lambda.
\end{equation}
\begin{figure}[H]
\centering
\includegraphics[scale=0.4]{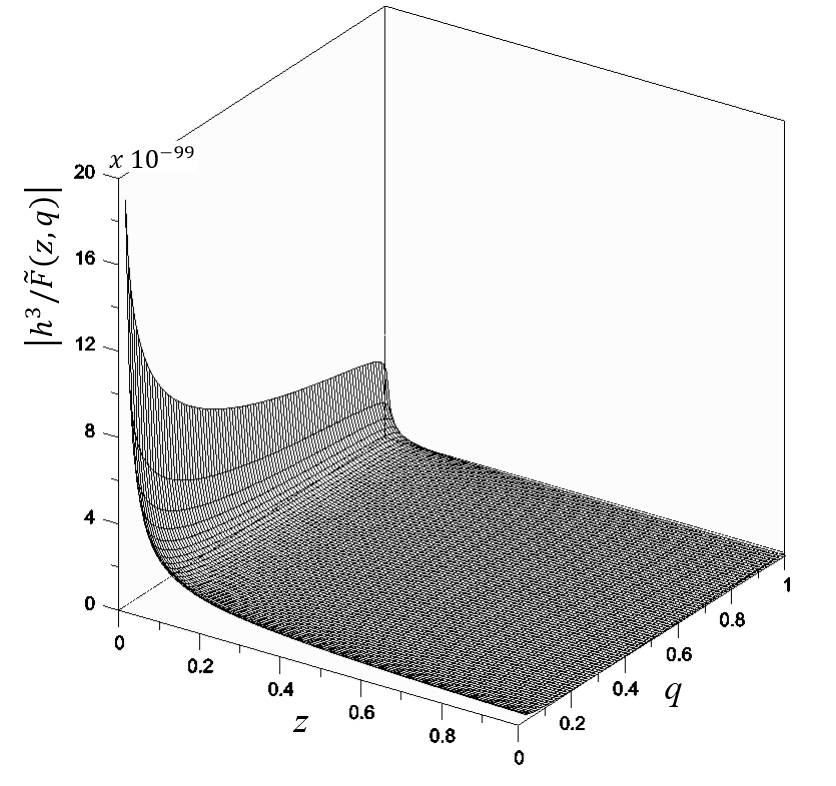} 
\caption{The $q$-deformed reduction factor $|{h^3}/{\widetilde{F}\left(z,q\right)}|$ as a function of $z$ for the case $q<1$.}
\label{fig: resim1}
\end{figure}
The equation (\ref{eq: dE_eq}) corresponds to generalised Einstein equations for
$q$-deformed fermion gas systems under high-temperature condition
and it contains the effective cosmological constant and the effective gravitational constant. As we mention in Introduction, this equation has certain components of modified Einstein
equations. Here $\alpha\left(z,q\right)^{-1}$ is a factor that carrying
the information of deformed fermion gas system. In order to compare the behaviours of the effective gravitational constant $G^{\left(q\right)}$ and the standard gravitational constant \textit{$G$}, in Fig. (\ref{fig: resim1}), the plot of the reduction factor $|{h^3}/{\widetilde{F}\left(z,q\right)}|$ is shown as a function of \textit{$z$} for the several values of the deformation parameter \textit{$q$} for the interval  \textit{$0<q<1$}. According to Fig. (\ref{fig: resim1}), the values of reduction factor decrease with the value of the deformation parameter \textit{$q$}. Moreover, this factor cause a decrease in the standard gravitational constant \textit{$G$}. Therefore, the standard gravitational constant \textit{$G$} is approximately \textit{$10^{-99}$} times smaller than the effective gravitational constant $G^{\left(q\right)}$.
On the other hand, Eq.(\ref{eq: dE_eq}) can be written as below,
\begin{equation}
R_{ab}-\frac{1}{2}g_{ab}R=8\pi G^{\left(q\right)}\left(T_{ab}+T_{ab}^{\Lambda}\right),
\end{equation}
where $T_{ab}^{\Lambda}$ can be defined as energy-momentum tensor
of vacuum and using the definition of $\Lambda^{\left(q\right)}$
and $G^{\left(q\right)}$ we can write following expression,
\begin{equation}
T_{ab}^{\Lambda}=-\frac{\Lambda^{\left(q\right)}}{8\pi G^{\left(q\right)}}g_{ab}=-\frac{\Lambda}{8\pi G}g_{ab}=-\left(\begin{array}{cc}
\rho_{\Lambda} & 0\\
0 & p_{\Lambda}g_{ij}
\end{array}\right),\label{eq: EM_tensor}
\end{equation}
and we obtain vacuum-energy density $\rho_{\Lambda}$ and vacuum pressure
$p_{\Lambda}$ as well-known form,
\begin{equation}
\rho_{\Lambda}=-p_{\Lambda}=\frac{\Lambda}{8\pi G}.
\end{equation}
This result shows that these quantities are not affected by the VPJC-type
fermionic $q$-deformation of the system. In the next section, the results obtained in this study will be given by comparing some studies in the literature.

\section{Conclusion}
This paper is related to the investigation of the effects of fermionic
\textit{$q$}-deformation on the Einstein field equations for high-temperature limit. We give
VPJC-type fermion oscillators algebra introduced in \cite{Viswanathan1992,Chaichian1993}.
We also represent some the high-temperature thermostatistical functions
of the deformed fermion gas model which previously studied in \cite{Algin2012}.
The deformed Einstein equations are obtained by applying Verlinde's entropic
gravity approach to the deformed entropy function. Also, the effective cosmological constant and the effective gravitational constant are found in terms of the factor $\alpha\left(z,q\right)^{-1}$ and, respectively, the standard cosmological constant and the standard gravitational constant. According to Fig. (\ref{fig: resim1}), the reduction factor $|{h^3}/{\widetilde{F}\left(z,q\right)}|$ decreases when the deformation parameter \textit{$q$} is increased. In this sense, we can say that the standard gravitational constant is \textit{$10^{-99}$} times smaller than the effective gravitational constant. 

Moreover, with the help of Eq.(\ref{eq: EM_tensor}), we obtained the vacuum-energy
density, which plays an important role to describe the cosmological
formation, in standard form as $\rho_{\Lambda}$=($\Lambda/8\pi G$) \cite{Cheng2005}. This result shows that the $q$-deformed system may not affect the evolution
of the Universe. The entropic gravity based on deformed
thermodynamical quantities provide an alternative way to describe
physical problems, for example, if we use different deformed thermodynamical
quantities we may obtain some contributions to the dark energy because
we know that there is a close relationship between the dark energy
and the cosmological constant.

According to Strominger's work in \cite{Strominger1993}, extremal
black holes obey deformed Bose or Fermi statistics. In this study, we consider them as deformed fermions and obtain a \textit{$q$}-deformed gravitational definition for them. In the limit \textit{$q\rightarrow1$}, the deformed fermion gas model reduces to standard fermion gas model. For this reason, the quantum black hole term vanishes because it only obeys the deformed statistics instead of the standard Bose or Fermi statistics. Therefore, \textit{$q$}-deformed Einstein equations do not reduces to standard Einstein equations when taken into account Strominger's proposal.

On the other hand, a parallel discussion was made by the works of Dil
with the use of different deformed systems \cite{Dil2015,dil2017}. In the Ref. \cite{Dil2015}, the extremal black holes were considered as deformed bosons and obtained deformed Einstein equations for them. In the Ref. \cite{dil2017}, there were found two-parameter deformed Einstein equations by assuming the extremal black holes as \textit{$(p,q)$}-deformed fermions. Moreover, in the Ref. \cite{sheykhi2012}, the modified Einstein equations were attained by considering the Debye correction to the equipartition law of energy in the framework entropic gravity scenario. As a consequence, the results in this study have different properties from these studies in the literature \cite{sheykhi2012,Dil2015,dil2017}.

The investigation of deformed Einstein equation by applying the low-temperature behaviour of the present deformed fermion model is a open problem to study in the near future.


\begin{thebibliography}{10}
\bibitem{hawking1971} S. W. Hawking, Phys. Rev. Letters \textbf{26},
1344 (1971). 

\bibitem{hawking1975} S. W. Hawking, Commun. Math. Phys. \textbf{43},
199 (1975).

\bibitem{bekenstein1973} J. D. Bekenstein, Phys. Rev. D \textbf{7},
2333 (1973).

\bibitem{jacobson1995} T. Jacobson, Phys. Rev. Lett. \textbf{75},
1260 (1995).

\bibitem{verlinde2011} E. Verlinde, JHEP \textbf{04}, 029 (2011).

\bibitem{mann2011} R. B. Mann, J. R. Mureika, Phys. Lett. B \textbf{703},
167 (2011).

\bibitem{li2012} M. Li, M. Rong-Xin, M. Jun, arXiv:hep-th/1207.0661v2.

\bibitem{padmanabhan2010} T. Padmanabhan, Rept. Prog. Phys. \textbf{73},
046901 (2010).

\bibitem{sheykhi2012} A. Sheykhi, S. K. Rezazadeh, J. Cosmol. Astropart.
Phys. \textbf{10}, 012 (2012).

\bibitem{ho2010} C. M. Ho, D. Minic, Y. J. Ng, Phys. Lett. B \textbf{693},
567 (2010).

\bibitem{ng2013} Y. J. Ng, Int. J. Mod. Phys. D \textbf{22}, 1342022
(2013).

\bibitem{edmonds2014} D. Edmonds, D. Farrah, C. M. Ho, D. Minic,
Y. J. Ng, T. Takeuchi, ApJ \textbf{793}, 41 (2014).

\bibitem{hendi2012} S. H. Hendi, A. Sheykhi, Int. J. Theor. Phys.
\textbf{51}, 1125 (2012).

\bibitem{Dil2015} E. Dil, Can. J. Phys. \textbf{93}, 1274 (2015).

\bibitem{dil2016B} E. Dil, E. Kolay, Adv. High Energy Phys. \textbf{2016},
3973706 (2016).

\bibitem{dil2017} E. Dil, Int. J. Mod. Phys. A \textbf{32}, 1750080
(2017). 

\bibitem{Strominger1993} A. Strominger, Phys. Rev. Lett.\textbf{
71}, 3397 (1993).

\bibitem{Gavrilik2011} A. M. Gavrilik, I. I. Kachurik, Y. U. Mishchenko,
J. Phys. A: Math. Theor. \textbf{44}, 475303 (2011).

\bibitem{Gavrilik2013} A. M. Gavrilik, Yu. A. Mishchenko, Ukr. J.
Phys. \textbf{58}, 1171 (2013).

\bibitem{Algin2013} A. Algin, E. Ilik, Phys. Lett. A \textbf{377},
1797 (2013).

\bibitem{Algin2015} A. Algin, D. Irk, G. Topcu, Phys. Rev. E \textbf{91},
062131 (2015). 

\bibitem{Cai2007} S. Cai, G. Su, J. Chen, J. Phys. A\textbf{ 40},
11245 (2007).

\bibitem{Mirza2011} B. Mirza, H. Mohammadzadeh, J. Phys. A: Math.
Theor. \textbf{44}, 475003 (2011).

\bibitem{Lavagno2000} A. Lavagno, P. Narayana Swamy, Phys. Rev. E
\textbf{61}, 1218 (2000).

\bibitem{Lavagno2002} A. Lavagno, P. Narayana Swamy, Phys. Rev. E
\textbf{65}, 036101 (2002).

\bibitem{Lavagno2010} A. Lavagno, P. Narayana Swamy, Found. Phys.
\textbf{40}, 814 (2010).

\bibitem{Marinho2012} A. A. Marinho, F. A. Brito, C. Chesman, Physica
A \textbf{391}, 3424 (2012).

\bibitem{Zeng2012} O. J. Zeng, Z. Chen, J. H. Yuan, Physica A \textbf{391},
563 (2012).

\bibitem{NarayanaSwamy2006a} P. Narayana Swamy, Int. J. Mod. Phys.
B \textbf{20}, 2537 (2006).

\bibitem{NarayanaSwamy2006b} P. Narayana Swamy, Eur. Phys. J. B \textbf{50},
291 (2006).

\bibitem{Sviratcheva2004} K. D. Sviratcheva, C. Bahri, A. I. Georgieva,
J. P. Draayer, Phys. Rev. Lett. \textbf{93}, 152501 (2004).

\bibitem{Bonatsos1992} D. Bonatsos, C. Daskaloyannis, Phys. Rev.
A \textbf{46}, 75 (1992).

\bibitem{Bonatsos1996} D. Bonatsos, C. Daskaloyannis, Mod. Phys.
Lett. B \textbf{10}, 1011 (1992).

\bibitem{Monterio1996} M. R. Monterio, L. M. C. S. Rodrigues, Phys.
Rev. Lett. \textbf{76}, 1098 (1996).

\bibitem{Parthasarathy1991} R. Parthasarathy, K. S. Viswanathan, J. Phys. A \textbf{24}, 613 (1991).

\bibitem{Viswanathan1992} K. S. Viswanathan, R. Parthasarathy, R.
Jagannathan, J. Phys. A \textbf{25}, L335 (1992).

\bibitem{Chaichian1993} M. Chaichian, R. G. Felipe, C. Montonen,
J. Phys. A \textbf{26}, 4017 (1993).

\bibitem{Algin2011} A. Algin, Int. J. Theor. Phys. \textbf{50}, 1554
(2011).

\bibitem{Algin2012} A. Algin, M. Senay, Phys. Rev. E \textbf{85},
041123 (2012).

\bibitem{wald1984} R. M. Wald, \textit{General Relativity}, (The
University of Chicago Press, Chicago, 1984).

\bibitem{Cheng2005} T. Cheng, \textit{Relativity, Gravitation, and Cosmology}, ( Oxford University Press, New York, 2005).
\end{thebibliography}
\end{document}